%Paper: hep-th/9512080
%From: gagne@physics.ucla.edu (Darius Gagne)
%Date: Mon, 11 Dec 1995 15:48:44 -0800
%Date (revised): Tue, 19 Dec 1995 16:53:16 -0800

\magnification=1200
\font\stepfour=cmr10 scaled\magstep3
\font\stepone=cmr10 scaled\magstep1

\hfuzz=12pt
\baselineskip=15pt
\def\makefootline{\baselineskip=36pt\line{\the\footline}}
\nopagenumbers

% slash.tex  --  some slashed characters for use in math mode
%
% "\slashsym<num><sym>" gives slashed version of <sym>; slash is
% shifted
% to right by <num>mu. Example: "\slashsym9A" looks about right.
\catcode`\*=11  % use * as ordinary char temporarily
\def\slashsym#1#2{\mathpalette{\sl*sh{#1}}{#2}}
\def\sl*sh#1#2#3{\ooalign{\setbox0=\hbox{$#2\not$}
                          $\hfil#2\mkern-24mu\mkern#1mu
                           \raise.15\ht0\box0\hfil$\cr
                          $#2#3$}}
\catcode`\*=12

\def\pslash{{\slashsym8p}}

\def\Aslash{{\slashsym9A}}

\def\Bslash{{\slashsym6B}}

\def\dslash{{\slashsym5\partial}}

\def\half{{1 \over 2}}

\rightline{UCLA/95/TEP/35}
\vskip .3in

\centerline{\bf\stepfour Worldline Path Integrals for Fermions }
\medskip
\centerline{{\bf\stepfour  with General Couplings }
\footnote{${}^\dagger$} {Research supported in part by National
Science  Foundation grant PHY-92-18990}}

\vskip 0.7in

\centerline{{\stepone Eric D'Hoker and Darius G. Gagn\'e}\footnote{${
\ }^*$}
{E-mail addresses : {\tt dhoker@physics.ucla.edu and
gagne@physics.ucla.edu}}}
\medskip
\centerline{\it Department of Physics and Astronomy}
\medskip
\centerline{\it University of California, Los Angeles}
\medskip
\centerline{\it Los Angeles, CA 90024, USA}

\vskip .8in

\centerline{\bf Abstract}

\vskip .2in

We derive a worldline path integral representation for the effective
action of a multiplet of Dirac fermions coupled to the most general
set of  matrix-valued scalar, pseudoscalar, vector, axial vector and
antisymmetric tensor background fields.  By representing internal
degrees of   freedom in terms of worldline fermions as well, we obtain
a formulation which   manifestly  exhibits chiral gauge invariance.

\vskip 0.5in

\centerline{\it Dedicated to the Memory of Shechao Feng}

\vfill
\eject

\footline={\hss\tenrm\folio\hss}

\noindent
{\bf 1. Introduction}

\bigskip

The worldline path integral reformulation of quantum field theory
produces an efficient alternative way of evaluating one loop
Feynman diagrams and effective actions.
The worldline formalism has a long history [1]; its close
connection with string theory was studied in [2] and its use in the
study of anomalies was demonstrated in [3].
The approach has known a revival of interest over
the past few years, once
it was realized that it gives rise to an improved method for
computing large numbers of complicated Feynman diagrams [4,5].

Most of the investigations, though, have concentrated on QED-type
diagrams with electrons and photons, and on QCD-type diagrams with
gluons, ghosts and vector-like quarks [4,5]; only   recently
has a stronger interest developed in effective
actions for fermions coupled to other fields as well. The case
of couplings to a single scalar and pseudoscalar field was
presented in [6], while the case of matrix-valued
scalar, pseudoscalar and vector gauge fields was treated
in [7]; the couplings to a single scalar,
pseudoscalar, Abelian vector and Abelian axial vector have been
recently discussed in [8], while the antisymmetric tensor coupling is
discussed in [9].

More specifically, in our preceding work in Ref. [7], we derived, in
a systematic way,  the worldline path integral representation of the
(one-loop)
effective action for a multiplet of Dirac fermions coupled to general
matrix-valued scalar, pseudoscalar, and vector gauge background
fields.
In a perturbative expansion in weak background fields, the real part
of  the Euclidean effective action generates Feynman graphs with an
even  number of $\gamma _5$-vertices while the imaginary part
generates graphs with an odd number of $\gamma   _5$-vertices.
Real and imaginary parts of the (Euclidean space-time) effective
action are reformulated in terms of worldline fermion integrals with
anti-periodic and periodic boundary conditions respectively.

In the present paper, we use and extend the technology developed in  [7]
to derive systematically the worldline path integral representation
for the effective action of a multiplet of Dirac fermions
coupled to the most general set of matrix-valued  scalar,  pseudoscalar,
vector, axial vector and antisymmetric tensor background fields.
The presence of both vector and axial vector couplings allows us to deal
also with chiral fermions, which is required in the study and computation of
Feynman amplitudes in electro-weak theory and in grand-unified models.
With the further presence of antisymmetric tensor background fields,
we complete
the list of all possible couplings of fermions to fields of spin less than
or equal
to one. Furthermore, when fermions are coupled to background
geometry, i.e. background gravity, couplings to antisymmetric fields naturally
appear through the appearance of the spin connection.

The effective action for fermions in the presence of this general set of
background fields is again obtained as the sum of real and   imaginary
parts, which are represented as path integrals over worldline
fermions
with anti-periodic and periodic boundary conditions respectively.
This sum over different boundary conditions is analogous to the
sum over spin structures familiar from superstring theory rules.

Throughout, we shall seek to maintain chiral (gauge) invariance
as much as is possible. The real part of the effective action
admits a completely
chiral invariant formulation, while the imaginary part -- which
gives rise, amongst other contributions,
to chiral anomalies -- admits a formulation that
involves an interpolation between chiral invariant Lagrangians,
but is not itself manifestly chiral invariant.
The imaginary part also requires an additional insertion operator
(analogous to the worldsheet supercurrent insertion in superstring
theory [10]), which converts the worldline fermion zero modes into a
space-time Levi-Civita $\varepsilon$ tensor.

The remainder of this paper is arranged as follows. In Sect. 2,
we express the fermion effective action in terms of a functional
determinant of the Dirac operator ${\cal O}$, suitably continued
to Euclidean space-time. In Sect. 3, its real part is reformulated
in terms of a path integral involving worldline
fermions over anti-periodic boundary conditions, and with
manifest chiral (gauge) invariance. In Sect. 4, its imaginary
part is re-expressed in terms of an integral over worldline
fermions involving period boundary conditions, but manifest
chiral (gauge) invariance is lost here.
In Sect. 5, we show how the internal degrees of freedom can
be represented by a new set of worldline anti-commuting degrees
of freedom. In this way, the final
worldline Lagrangians may be expressed as $c$-numbers.

In Appendix A, we present an alternative reformulation of the
imaginary
part of the fermion effective action, in terms of a worldline path
integral
which is manifestly invariant under vector-like gauge
transformations, and is
closer in spirit to our construction in [7].

\vskip 0.3in

\noindent
{\bf 2. Effective Action}

\bigskip

We study the effective action of a multiplet of $N$ Dirac fermions
coupled to the an arbitrary set of matrix-valued scalar $\Phi(x)$,
pseudoscalar  $\Pi(x)$, vector  $A _\mu (x)$, axial vector $B _\mu (x)$
and antisymmetric   tensor
$K_{\mu \nu} (x)$ background fields.  The most general, CPT
\vfill\break
\noindent
invariant, classical fermion action is given by
\footnote{*}{
Here, the Minkowski space-time metric $\eta$ has signature $
(+~-~-~-)$ and
we make use of the standard conventions
$\{\gamma^{\mu},\gamma^{\nu}\} = 2\eta^{\mu\nu}$,
${\gamma^5} = i\gamma^0\gamma^1\gamma^2\gamma^3 $, and
$\varepsilon_{0123} = -\varepsilon^{0123} = 1$.}
$$
S[{\bar \Psi} ,\Phi ,\Pi ,A,B,K,\Psi ]
=
\int\limits d^4x~
{\bar \Psi}^I \bigl [i\dslash - \Phi + i\gamma^5\Pi + \Aslash +
\gamma^5\Bslash
  + i\gamma^{\mu}\gamma^{\nu}K_{\mu\nu}]^{IJ} \Psi^J \; .
\eqno (2.1)
$$
The background fields are Hermitian; coupling constants have been
absorbed
into the definition of the background fields; the superscripts, $I$
and
$J$, refer to the  internal quantum numbers of the fermion multiplet
and of the
matrix-valued background fields.    We might expect the presence of a
further
coupling term in (2.1) of the form $\gamma ^5 \gamma ^\mu \gamma^\nu
K'_{\mu
\nu}$ with an additional background field $K'_{\mu \nu}$. However,
this term
need
not be included, as it is simply related to the $K_{\mu \nu}$ term
already
present in (2.1). This fact can be seen from the identity
$$
i\gamma^{\mu}\gamma^{\nu}K_{\mu\nu}
=
\gamma^5\gamma^{\mu}\gamma^{\nu}\tilde{K}_{\mu\nu} ~,
\qquad {\rm with} \qquad
\tilde{K}_{\mu\nu} \equiv {1\over 2} \varepsilon_{\mu \nu \rho
\sigma}
					K^{\rho \sigma} ~,
\eqno (2.2)
$$
where $\tilde K$ is the dual of $K$.
The largest internal unitary symmetry group of  the action
$S$ is $U(N)_L \times U(N)_R$, obtained
e.g. for zero background fields; more generally, we shall assume that
a
subgroup $G_L \times G_R$ of $U(N)_L\times U(N)_R$ is gauged. It is
convenient to introduce chiral fields, which transform simply under
$G_L \times G_R$, as follows
$$
\Psi ^{{L\atop R}}  =  {1\over 2}(1 \mp \gamma^5)\Psi,
\qquad
A^{{L \atop R}} =  A\pm B,
\qquad
H\ = \Phi -i \Pi,
\qquad
K^s =  K -i \tilde K ~.
\eqno (2.3)
$$
The complex antisymmetric tensor field $K^s _{\mu \nu}$ is
``self-dual"
with $\tilde K^s_{\mu \nu} = i K^s _{\mu \nu}$ and provides a
decomposition
of an
arbitrary real antisymmetric tensor into complex conjugate self-dual
and
anti-self-dual components : $2K = K^s + (K^s)^\dagger$.
Under $G_L \times G_R$,
$\Psi ^L$ and $\Psi^R$ transform as
$N$-dimensional representations $T_L\otimes 1$ and $1 \otimes T_R$
respectively,
while $H$ and $K^s$ transform as the $N \times N$ dimensional
representation
$T_L \otimes T_R ^*$.

The effective action $W[\Phi,\Pi,A,B,K]$ for fermions in the presence
of
the above background fields may be defined by a functional
determinant in
Minkowski space-time
$$
iW[\Phi,\Pi,A,B,K] = \log{\rm Det}\,
i[i \dslash  - \Phi + i\gamma^5\Pi + \Aslash + \gamma^5\Bslash
+ i\gamma^{\mu}\gamma^{\nu}K_{\mu\nu}]\; .
\eqno (2.4)
$$
To utilize heat-kernel methods we first analytically continue the
theory to
Euclidean space as in [7].  Although the $\gamma$-matrices are
unaffected
by the continuation, it is useful to change notation to Hermitian
generators of
an Euclidean Clifford algebra, $(\gamma_E)_j \equiv i\gamma_j$,
$(\gamma_E)_4 \equiv \gamma_0$, and $(\gamma_E)_5 \equiv \gamma_5$,
satisfying $\{(\gamma_E)_a,(\gamma_E)_b\} = 2\delta_{ab}$  with
$a,b = \mu,5$.
{}From the Wick-rotation,  $t\rightarrow -it$, it follows that
$$
\dslash \rightarrow i\dslash_E ~,
\quad
\Aslash \rightarrow i\Aslash_E ~,
\quad
\Bslash \rightarrow i\Bslash_E ~,
\quad
\gamma^{\mu}\gamma^{\nu}K_{\mu\nu}
\rightarrow
-(\gamma_E)_{\mu}(\gamma_E)_{\nu}K_{E\mu\nu} \; .
\eqno (2.5)
$$
The expression (2.4) for the fermion effective action translates
into the following Euclidean space-time functional determinant form :
$$
-W[\Phi , \Pi,A,B,K] = \log{\rm Det}\, [{\cal O}] \; ,
\eqno (2.6)
$$
where the operator ${\cal O}$ is defined by
\footnote{${}^\dagger$}{
Henceforth, space-time is taken to be Euclidean and the subscript $E$
is dropped. Also, $ p = -i\partial$ is the momentum conjugate to
$x$,  both of which are Hermitian operators.}
$$
{\cal O} \equiv \pslash  - i\Phi (x) - \gamma_5\Pi (x) - \Aslash (x)
- \gamma_5\Bslash (x) + \gamma_{\mu}\gamma_\nu K_{\mu\nu}(x) \; .
\eqno (2.7)
$$
As in [7], it is convenient to split the effective action into its
real and imaginary parts : $ W= W_{\Re}+  iW_{\Im}$. Under parity
transformations, both contributions are separately invariant. A
perturbative expansion in weak fields reveals that graphs with an
even number of $\gamma_5$-vertices are real and contribute to
$W_{\Re}$ while  those with an odd number of
$\gamma_5$-vertices are imaginary and contribute to $W_{\Im}$.

\bigskip
\bigskip

\noindent
{\bf 3. Worldline Path Integral for the Real Part of the Effective
Action}

\bigskip

We propose to obtain a worldline path integral formulation for
$W_\Re$ with manifest chiral gauge invariance, and a worldline
Lagrangian that is even in worldline fermions. First, we double the
fermion system in terms of a Hermitian operator $\Sigma$, as in [7]
:
$$
W_{\Re} = -{ 1 \over 2} \ln {\rm Det} [{\cal O} ^\dagger
{\cal O}]= -{1\over 2}\ln {\rm Det}\, [\Sigma],
\qquad \quad
{\Sigma} \equiv  \left(\matrix{ 0                      & {\cal O}
\cr
                              {\cal O}^{\dagger} & 0
\cr}
\right) .
\eqno (3.1)
$$
We introduce six-dimensional Hermitian Euclidean $\Gamma_{\rm A}$
matrices,   satisfying $\{\Gamma_{\rm A} , \Gamma_{\rm B}\} =
2\delta_{\rm AB}\, {\rm   I}_{8}$, for ${\rm A,B} = 1,...,6$, and
choose a basis where
\footnote{*}{We take a basis where
$\gamma _5 = \pmatrix{ {\rm I}_2 & 0 \cr
0 & -{\rm I}_2 \cr}$. ${\rm I}_d$ denotes the $d\times d$ identity matrix.}
$$
\Gamma_{\mu} \equiv \left(\matrix{ 0            & \gamma_{\mu} \cr
                              \gamma_{\mu} & 0            \cr}\right)
{}~ ,
\qquad
\Gamma_5     \equiv \left(\matrix{ 0            & \gamma_5     \cr
                              \gamma_5     & 0            \cr}\right)
{}~ ,
\qquad
\Gamma_6     \equiv \left(\matrix{ 0            & i{\rm I}_4     \cr
                                 -i{\rm I}_4    & 0
\cr}\right) ~ .
\eqno (3.2)
$$
For later use, we also introduce the Hermitian matrix $\Gamma _7$
with
$$
\Gamma_7 \equiv -i\prod_{{\rm A}=1}^6\Gamma_{\rm A}
              = \pmatrix{ {\rm I}_4  &  0          \cr
                          0          &  -{\rm I}_4 \cr}\; ,
\qquad
\{ \Gamma _{\rm A}, \Gamma _7 \} =0 ~~ {\rm for}~~ {\rm A}=1, \cdots
,6.
\eqno (3.3)
$$
The operator $\Sigma$ is simply expressed in terms of
$\Gamma$-matrices, in analogy with [7] :
$$
\Sigma = \Gamma _\mu (p_\mu - A_\mu) - \Gamma _6 \Phi - \Gamma _5 \Pi
-i \Gamma _\mu \Gamma _5 \Gamma _6 B_\mu -i\Gamma _\mu \Gamma _\nu
\Gamma _6 K_{\mu \nu} \; .
\eqno (3.4)
$$
The form of $\Sigma$ in (3.4) does not exhibit manifest chiral
covariance, since the vector and axial gauge fields multiply
independent  $\Gamma$-matrices.
To circumvent the problem, we notice that $i \Gamma _5 \Gamma _6$
has eigenvalues 1 and $-1$, and that within the corresponding
eigenspaces, gauge
fields couple in a manifestly chiral way through $A\pm B$. Thus, to
render the chiral covariance manifest, we   re-label
$i\Gamma _5 \Gamma _6$ as an internal quantum number. This reduces
the Clifford representation in (3.4) to a four-dimensional one, but
doubles the dimension   of the internal quantum numbers. This
re-interpretation is equivalent to  re-labeling all right-handed Weyl
fermions as charge conjugates of left-handed Weyl   fermions.

To render this re-labeling more explicit, it is convenient to work in a
new $\Gamma $-matrix basis, with matrices denoted by $M^{-1}\Gamma M$
and arranged so that  $M^{-1}i\Gamma _5 \Gamma _6 M$ is simple :
$$
M^{-1} i\Gamma   _5 \Gamma  _6 M =
\left(\matrix{ {\rm I}_4                &  0 \cr
               0 &  -{\rm I}_4        \cr}\right),
\qquad \qquad
M=\left(\matrix{
{\rm I}_2  &   0 &   0  &  0     \cr
0    &   0 &   0  &  {\rm I}_2   \cr
0    &   0 & {\rm I}_2  &  0     \cr
0    & {\rm I}_2 &   0  &  0     \cr
}\right).
\eqno (3.5)
$$
In this basis, the operator $\Sigma$ takes on a manifestly chiral
invariant form, as can easily be seen by expressing it in terms of the
chiral fields   of (2.3) :
$$
M^{-1}\Sigma M =
\left(\matrix{
\gamma _\mu (p_\mu - A^L_\mu)
&  &
\gamma_5(-iH + \half\gamma_\mu\gamma_\nu K^s_{\mu\nu})
\cr
&& \cr
\gamma_5(iH^\dagger  - \half\gamma_\mu\gamma_\nu (K^s _{\mu\nu})
^\dagger)    &  &
\gamma _\mu (p_\mu  - A^R_\mu)
\cr}\right) \; .
\eqno (3.6)
$$
To represent the operator $\Sigma$ in terms of a worldline path
integral,
we must find a representation of $\gamma _\mu$ and $\gamma _5$ in
terms of
coherent states. But, we know from the analysis of [7] that this
representation cannot be achieved in terms of $4\times 4$ matrices.
Instead, we shall double up the fermion system once more to achieve
a good coherent state representation of the Clifford algebra.

The simplest way to achieve this is by replacing the $4\times 4$
Clifford matrices $\gamma $ everywhere in (3.6) by the $8\times 8$
Clifford   matrices $\Gamma $, as defined in (3.2). The new operator,
thus obtained,   will be denoted by $\tilde \Sigma $, and we have
$$
W_{\Re} = -{1\over 4}\ln  {\rm Det}\, [ \tilde\Sigma ] \; ,
\eqno (3.7)
$$
$$
\tilde\Sigma =   \Gamma_\mu (p_\mu - {\cal A}_\mu )
- \Gamma_5 {\cal H}
- {i \over 2} \Gamma_\mu\Gamma_\nu\Gamma_5 {\cal K}_{\mu\nu} ~ .
\eqno (3.8)
$$
The $2N \times 2N$ Hermitian background fields are
defined in terms of the chiral fields of (2.3)
$$
{\cal A}_\mu \equiv \left(\matrix{A^L_\mu & 0       \cr
                                  0       & A^R_\mu \cr}\right) ~ ,
\qquad
{\cal H}    \equiv \left(\matrix{0          & iH \cr
                                  -iH^\dagger & 0 \cr}\right) ~ ,
\qquad
{\cal K}_{\mu\nu} \equiv
          \left(\matrix{0                   & iK^s_{\mu\nu} \cr
                         -i(K^s_{\mu\nu})^\dagger & 0
\cr}\right) \; .
\eqno (3.9)
$$
Finally, the operator $\tilde\Sigma$ is manifestly chiral gauge
covariant, which will guarantee that the real part of the Euclidean
effective action is manifestly invariant.

To obtain a well-defined heat-kernel regularization of the
determinant
defining $W_\Re$, we replace the self-adjoint operator $\tilde
\Sigma$
by its square, which is automatically positive, and find
\footnote{${}^\dagger$}{We denote by Tr functional traces of
operators
and reserve tr for traces of finite dimensional internal matrices.}
$$
W_{\Re} = -{1\over 8} \ln {\rm Det} \,  [\tilde\Sigma^2 ]
 = {1\over 8}\int_0^{\infty}{dT\over T}
                  {\rm Tr}\, e^{-{{\cal E}\over
2}T\tilde\Sigma^2}\;\;,
\eqno (3.10)
$$
where the operator $\tilde\Sigma ^2$ works out to be
$$
\eqalign{
{\tilde\Sigma}^2
&
= (p - {\cal A})^2 + {\cal H}^2
+ {1 \over 2}{\cal K}_{\mu\nu}{\cal  K}_{\mu\nu}
+ {i \over 2} \Gamma_{\mu} \Gamma_{\nu}
  \bigl ( {\cal F}_{\mu\nu} +
  \{ {\cal H},{\cal K}_{\mu\nu} \} +
  i [ {\cal K}_{\mu\rho} , {\cal K}_{\rho\nu}] \bigr )  \cr
&
+ i\Gamma _\mu \Gamma _5 \bigl ( {\cal D}_\mu  {\cal H}+
  \{ p_\nu - {\cal A}_\nu , {\cal K}_{\mu\nu} \} \bigr )
- \half \Gamma_{\mu\rho\sigma}\Gamma_5{\cal D}_\mu
 {\cal K}_{\rho\sigma}
- { 1 \over 4} \Gamma_{\mu\nu\rho\sigma}
{\cal K}_{\mu\nu} {\cal K}_{\rho\sigma} \; . \cr}
\eqno (3.11)
$$
Here, $\Gamma _{{\rm A}_1 \cdots {\rm A}_k} \equiv \Gamma _{[{\rm
A}_1} \cdots \Gamma _{{\rm A}_k]}$ denotes the anti-symmetrized
product of $k$ $\Gamma$ matrices.   Covariant derivatives and fields
strengths have been defined relative to the   field ${\cal A}_\mu$ :
$$
{\cal F}_{\mu\nu} = \partial _\mu {\cal A}_\nu - \partial _\nu {\cal
A}_\mu
-i [{\cal A}_\mu , {\cal A}_\nu ],
\qquad
{\cal D}_{\mu} {\cal H} = \partial _\mu {\cal H}  - i [{\cal A}_\mu ,
{\cal
H} ]
\eqno (3.12)
$$
and similarly for ${\cal D}_\mu {\cal K}_{\rho \sigma}$.
We have now achieved a second order formulation of the fermion
determinant
problem that manifestly exhibits the chiral gauge symmetries of the
original
Lagrangian while only involving terms that are even in the number of
$\Gamma _a $ matrices $a=1,\cdots, 5$, and it remains to derive a
worldline
path integral formulation for it.

In a preceding paper [7], we constructed a worldline fermion
reformulation of
the
Clifford algebra with the help of the coherent state formalism. We
showed
that under this correspondence, $\Gamma_{\rm A}\Gamma _{\rm B}
\rightarrow
2\psi_{\rm A}\psi_{\rm B}$, when A and B are different; analogously,
it is equally easy to show
$\Gamma_{\rm A}\Gamma_{\rm B}\Gamma_{\rm C}\Gamma_{\rm D} \rightarrow
4\psi_{\rm A}\psi_{\rm B}\psi_{\rm C}\psi_{\rm D}$, when A, B, C and
D are all
different. By inspection of (3.11), it is clear that these are the
only
Clifford multi-linears needed to express $W_\Re$ as a path integral
involving
worldline fermions. Since each term in $\tilde\Sigma^2$ is even in
powers of
$\Gamma$-matrices, we are guaranteed that our final worldline
Lagrangian will
have even  Grassmann grading as shown in [7].

Denoting the  dependence of the operator $\tilde \Sigma ^2 $ on
momentum, background fields and  the matrices $\Gamma_a$, $a =
1,\cdots, 5$, by $\tilde\Sigma^2 ( p, {\cal A}, {\cal H}, {\cal K},
\Gamma_a )$, the   trace of the exponential in (3.10) may easily be
written with the help of   the formalism of [7] :
$$
{\rm Tr}\, e^{-{{\cal E}\over 2}T{\tilde\Sigma}^2}
 = \int {\cal D}p{\cal D}x      \int \limits _{\rm AP} {\cal D}\psi
     ~{\rm tr}~{\cal P} e ^{\int_0^Td\tau\Bigl[ i\dot{x}_\mu
p_\mu       - {1\over 2}\psi_a\dot{\psi}_a
     - {{\cal E}\over 2}
\tilde\Sigma^2(p,{\cal A},{\cal H},{\cal K}, \sqrt{2}\psi_a)
\Bigr] } \; .
\eqno (3.13)
$$
Here, ${\cal P}$ and ${\rm tr}$ denotes path ordering of and tracing
over
internal representation matrices. In Sect. 5, we shall show that the
trace may be recast in terms of worldline Grassmann integrations as
well.  The   boundary conditions on the worldline loop are periodic for
$x$, while   anti-periodic (AP) for $\psi_a$. Notice that $\psi_6$ does
not enter in $\tilde\Sigma^2$   and may be integrated out of (3.13);
thus, the measure reduces to ${\cal D}\psi = {\cal D}\psi_\mu{\cal
D}\psi_5$.

The terms in the path integral which involve the momentum function
$p$ may be rearranged by completing the square in $p$. Using the
results   of Appendix A in [7], the momentum may then be decoupled from
all fields    by a suitable shift
\footnote{${}^\dagger$}
{Since the shift here involves worldline fermions it must actually be performed
at the level of the $\Gamma$-matrices.}
and yields a simple field-independent
normalization   factor ${\cal N}$.
$$
{\cal N} = \int {\cal D}p e^{- {{\cal E}\over 2} \int _0 ^T d \tau ~
p^2(\tau)}
\eqno (3.14)
$$
As a result, we find the following path integral representation for
the real part
$W_\Re$ of the effective action
$$
W_{\Re} = {1\over 8}\int_0^{\infty}{dT\over T}
                        {\cal N}\int{\cal D}x\int
                        \limits_{\rm AP}{\cal D}\psi ~
 {\rm tr} \, {\cal P} e^{-\int_0^Td\tau {\cal L}(\tau )}\; .
\eqno(3.15)
$$
The worldline Lagrangian, ${\cal L}(\tau)$, is given by
$$
\eqalign{
              {\cal L}(\tau)
	   &= {\dot{x}^2\over 2{\cal E}}
            + {1\over 2}\psi_a\dot\psi_a
            - i\dot{x}_\mu {\cal A}_\mu
            + {{\cal E}\over 2}{\cal H}^2
            - {{\cal E}\over 4}{\cal K}_{\mu\nu}{\cal K}_{\mu\nu}
            + i\psi_{\mu}\psi_5 \bigl ( {\cal E}{\cal D}_{\mu}{\cal
H}  +
              2i {\dot x}_\nu {\cal K}_{\mu\nu} \bigr )\cr
	   &
            + {i \over 2}{\cal E}\psi_{\mu}\psi_{\nu} \bigl (
                   {\cal F}_{\mu\nu} +
                   \{ {\cal H}, {\cal K}_{\mu\nu} \} \bigr)
            - {\cal E}\psi_{\mu} \psi_\nu \psi_\rho \bigl (
                   \psi_5{\cal D}_\mu {\cal K}_{\nu \rho} + \half
                   \psi _\sigma  {\cal K}_{\mu\nu} {\cal K}_{\rho
                   \sigma} \bigr ) \; . \cr}
\eqno (3.16)
$$
 The path integral of (3.15-16) is manifestly chiral gauge
invariant.   When ${\cal H}= {\cal K}_{\mu \nu}=0$, the fermion
multiplet is   massless, and
the worldline Lagrangian in (3.16) correctly reduces to the standard
form [5,7]
for a $2N$-component multiplet of massless fermions coupled to a
background
non-Abelian {\it vector gauge field}, ${\cal A}$.

The perturbative expansion of the path integral (3.15) proceeds by
decomposing $x(\tau) = x^\prime (\tau) + x^o$, where the zero mode
$x^o$ is constant and $ x^\prime $ is orthogonal to constants. The
$x'$ and $\psi _{\rm   A}$ propagators are :
$$
\eqalign{
\langle x'(\tau _1) x'(\tau _2) \rangle
& = {\cal E}{(\tau _1 - \tau _2)^2\over 2T} -
            {{\cal E} \over 2} \vert\tau _1 - \tau _2\vert +
            \;\;\hbox{constant}  \cr
\langle \psi _{{\rm A}_1} (\tau _1 ) \psi _{{\rm A}_2} (\tau _2)
\rangle
& = \half \delta _{{\rm A}_1 {\rm A}_2} ~ {\rm sign }(\tau _1 - \tau
_2) \cr}
\; .
\eqno (3.17)
$$
The normalization of the ${\cal D}x$ integral produces
an extra overall factor of $1/(2\pi {\cal E} T)^2$, while the
fermion integral produces an extra overall normalization factor of 8,
accounting for the dimensionality of the Clifford algebra.

\vskip 0.3in

\noindent
{\bf 4. Worldline Path Integral for the Imaginary Part of the
Effective Action}

\bigskip

In this Section, we propose to obtain a worldline path integral
formulation
for the imaginary part of the effective action $W_\Im$.  While it is
not
possible to preserve manifest chiral invariance for $W_\Im$, (which
generates,
amongst other things, the chiral anomaly), we shall seek a worldline
Lagrangian
that is as close as possible to the one for $W_\Re$, with as much
manifest
chiral symmetry as possible. First, we double the fermion system in
terms of a
{\it non}-Hermitian operator $\Omega$, as in [7] :
$$
W_{\Im} = -{\rm arg}{\rm Det}[{\cal O}]
        = -{1\over 2}{\rm arg}{\rm Det}\, [ \Omega ] ~,
\qquad
\Omega \equiv  \left(\matrix{ 0                      & {\cal O}  \cr
                              {\cal O} & 0
\cr}
\right) \; .
\eqno (4.1)
$$
The matrix $\Omega$ is simply represented in terms of the $\Gamma$-matrices of
(3.2-3) and we have
$$
\eqalign{
\Omega
 &
=  \Gamma _\mu (p_\mu - A_\mu) - \Gamma _7 \Gamma _6 \Phi - \Gamma  _5
     \Pi - i \Gamma _7\Gamma _\mu \Gamma _5 \Gamma _6  B_\mu -i \Gamma
     _7 \Gamma _\mu \Gamma _\nu \Gamma _6  K_{\mu \nu} \cr
&
= \half (\Sigma - \Gamma _6 \Sigma \Gamma _6)
  + \half \Gamma _7 (\Sigma + \Gamma _6 \Sigma \Gamma _6 ) \; .
\cr}
\eqno (4.2)
$$
Next, as in Sect. 3, we need to re-interpret $i \Gamma _5 \Gamma _6$ as
an internal quantum number, which will reduce the dimension of
the Clifford matrices back to $4\times 4$, and which will
double the dimension of the internal degrees of freedom. Then, as in
Sect. 3, we shall have to double the system once more to obtain a good
coherent state representation for the Clifford algebra. In the
present case of the operator $\Omega$, this procedure is
complicated  by the presence of $\Gamma _7$, and some additional
care is needed here.

To exhibit these operations most clearly, it turns out to be most
convenient to double the system first, and then transform to a basis
where the $\pm 1$ eigenvalues of the doubled $i\Gamma _5 \Gamma _6$ are
arranged in $8 \times 8$ blocks. This may be achieved in the following
way
\footnote{${}\dagger$}{This same procedure could have been used for obtaining
$\tilde{\Sigma}$ in the previous section but as we showed it was possible to
obtain $\tilde{\Sigma}$ in an equivalent but more direct way.}
$$
\tilde \Omega = \tilde M^{-1}
\pmatrix{ M^{-1}\Omega M & 0               \cr
                         &                 \cr
               0         & M^{-1} \Omega  M \cr} \tilde M \; ,
\qquad\qquad
\tilde M = \pmatrix{ {\rm I}_4  &   0   &   0   &   0   \cr
                       0   &   0   & {\rm I}_4  &   0   \cr
                       0   &  {\rm I}_4  &   0   &   0   \cr
                       0   &   0   &   0   &  {\rm I}_4  \cr} \; .
\eqno (4.3)
$$
Once this form is obtained, we may re-label the eigenvalues of the
doubled $i\Gamma _5 \Gamma _6$ as internal quantum numbers, and
re-express the operator $\tilde \Omega$ in terms of the $8\times 8$
$\Gamma$-matrices, defined in (3.2) and the chiral fields defined in
(3.9). The result is conveniently re-expressed in terms of the
operator $\tilde \Sigma$ of (3.8)
$$
\tilde \Omega = \half (\tilde \Sigma - \tilde \Sigma ^c) i\Gamma _6
\Gamma _7 +i \half \Gamma _5 \Gamma _6 \Gamma _7 \chi (\tilde \Sigma +
\tilde \Sigma ^c) i\Gamma _6 \Gamma _7 \; .
\eqno (4.4)
$$
Here, the operator $\tilde \Sigma ^c$ stands for $\tilde \Sigma$ of
(3.8), in which the fields ${\cal A}$, ${\cal H}$ and ${\cal K}$ have
been replaced by their chiral conjugates. This
transformation amounts to letting $A^{L,R}_\mu \mapsto A^{R,L}_\mu$,
$H \mapsto -H^\dagger$ and $K^s_{\mu \nu} \mapsto -(K^s_{\mu \nu})^\dagger$
and yields the
following fields
$$
{\cal A}_\mu ^c \equiv \left(\matrix{A^R_\mu & 0       \cr
                                  0       & A^L_\mu \cr}\right)  ,
\qquad
{\cal H} ^c    \equiv \left(\matrix{0          & -iH^\dagger \cr
                                  iH & 0 \cr}\right)  ,
\qquad
{\cal K}_{\mu\nu} ^c \equiv
          \left(\matrix{0                   & -i
(K^s_{\mu\nu})^\dagger
\cr
                         iK^s_{\mu\nu} & 0
\cr}\right)  .
\eqno (4.5)
$$
The matrix $\chi$ acts only on internal degrees of freedom, and
in this basis is given by
$$
\chi = \pmatrix{{\rm I}_N & 0    \cr
                  0  & -{\rm I}_N \cr}\; .
\eqno (4.6)
$$
We are now in a position to evaluate the functional determinant of
(4.1). As in [7], we shall make use of the heat-kernel
regularization for the arg-function,
$$
W_\Im
  =  -{1 \over 8} {\rm arg~Det} [\tilde \Omega ^2]
  = {i{\cal E}\over 64}
    \int_{-1}^1 \!\!\! d\alpha ~\int_0^\infty \!\!\! dT \,
    {\rm tr}\bigl \{
       (\tilde\Omega^2 - \tilde\Omega^{\dagger 2})
       e^{-{{\cal E}\over 2}T \tilde \Sigma _{(\alpha)} ^2
               } \bigr{\} } \; .
\eqno (4.7)
$$
The new operator $\tilde \Sigma _{(\alpha )} ^2$ depends on the
parameter $\alpha$, and is defined by
$$
\tilde \Sigma _{(\alpha)} ^2
=  {1 \over 4}(\tilde\Omega +
\tilde\Omega^{\dagger})^2
                   +{\alpha \over 2}[\tilde\Omega, \tilde\Omega
^\dagger]
                   -{\alpha ^2\over 4}
(\tilde\Omega - \tilde\Omega ^\dagger)^2
\eqno (4.8)
$$
To compute $\tilde \Sigma _{(\alpha )}^2$, it is helpful to  observe
that
$$
\tilde \Omega ^\dagger = \half (\tilde \Sigma - \tilde \Sigma ^c)
i\Gamma _6 \Gamma _7 -i \half \Gamma _5 \Gamma _6 \Gamma _7 \chi
(\tilde \Sigma + \tilde \Sigma ^c) i \Gamma _6 \Gamma _7\; .
\eqno (4.9)
$$
Furthermore, as explained in [7], due to the symmetry of the
$\alpha$-integration, the factor of $i\Gamma _5 \Gamma _6\Gamma_7
\chi$, which appears in the   commutator
$[\tilde \Omega, \tilde \Omega ^\dagger ]$, is immaterial and may be
omitted in the calculation of $\tilde \Sigma _{(\alpha )}^2$.
Finally, the overall factor of $i\Gamma _6 \Gamma _7$ cancels
out of the square in $\tilde \Sigma _{(\alpha )}^2$. Putting all
together, we discover that $\tilde \Sigma _{(\alpha )}^2$ in the
integration in (4.7), can be re-expressed in terms of
$\tilde \Sigma$ of (3.8), but where the  fields $\Phi$,
$B$ and $K$ have been rescaled by a factor of $\alpha$ :
$$
\tilde \Sigma _{(\alpha)}^2
= \tilde\Sigma^2 \big \vert_{\Phi  \rightarrow \alpha
\Phi ~,~
                      B     \rightarrow \alpha B   ~,~
                      K \rightarrow \alpha K }
                      \; .
\eqno (4.10)
$$
Notice that for   $\alpha =1$, we recover the original fields, and
the original operator $\tilde \Sigma ^2$ of (3.11), while for $\alpha
=-1$, we   recover all fields with chirality reversed, and the
operator $(\tilde \Sigma ^c)^2$, defined through (4.5).

Next, we derive an expression for the insertion operator $(\tilde
\Omega ^2 -\tilde \Omega ^{\dagger 2})$ in (4.7). From the structure
of $\tilde \Omega$ in (4.4), it appears natural to break up the
operator as
$$
\eqalign{
\tilde\Omega^2 - \tilde\Omega^{\dagger 2}
& =
i  ~\Gamma_7 \chi ~ \omega(p,{\cal A},{\cal H},{\cal K},\Gamma_a)
\Gamma_6 ~, \cr
\omega
& =
\Gamma _5 [\tilde \Sigma, \tilde \Sigma ^c] \; .
\cr}
\eqno (4.11)
$$
The reduced insertion operator $\omega$ may easily be worked out in
terms of the chiral fields of (3.9), and we find
$$
\eqalign{
\omega
&=
 \Gamma_\mu \bigl ( i\{  {\cal D}^c_\mu , {\cal H}\}
                     - i \{ {\cal D}_\mu , {\cal H}^c\}
                     + \{ {\cal D}^c_\nu ,{\cal K}_{\mu\nu}\}
                     - \{ {\cal D}_\nu ,{\cal K}^c_{\mu\nu} \} \bigr  )
			-  {1\over 4}\Gamma_{\mu\nu\rho\sigma} \Gamma_5
                     [{\cal K}_{\mu\nu},{\cal K}^c_{\rho\sigma}]
\cr
& 		+\Gamma_5   \bigl ( -[{\cal D}_\mu , {\cal D}^c_\mu ]
                     +  [{\cal H} , {\cal H}^c ]
       		+ \half [{\cal K}_{\mu\nu},{\cal K}^c_{\mu\nu}]\bigr)
  + \half \Gamma_{\mu\nu\rho} \bigl ( {\cal D}_\mu {\cal K}^c_{\nu\rho}
		 - {\cal D}^c_\mu {\cal K}_{\nu\rho} \bigr )
\cr
&+ \Gamma_{\mu\nu}  \Gamma_5  \bigl (
            - \{ {\cal D}_\mu , {\cal D}^c_\nu \}
             + {i \over 2} [{\cal H},{\cal K}^c_{\mu\nu} ]
             -  {i \over 2} [{\cal H}^c,{\cal K}_{\mu\nu} ]
             -  \{ {\cal K}_{\mu\rho},{\cal K}^c_{\rho\nu} \}\bigr ) \; .
\cr}
\eqno (4.12)
$$
The chiral fields ${\cal A}$, ${\cal H}$ and ${\cal K}$ were defined
in (3.9), while their charge conjugates are given in (4.5). Covariant
derivatives ${\cal D}_\mu$ are taken with respect to ${\cal A}$,
while ${\cal D}_\mu ^c$ is taken with respect to ${\cal A}^c$.

 The heat-kernel representation of (4.7), with the above results giving
$\tilde \Sigma _{(\alpha )}^2$ and the insertion $\tilde
\Omega ^2 - \tilde \Omega ^{\dagger 2}$ in terms of
$\Gamma$-matrices, allows for an immediate coherent state
reformulation. As we have shown in [7],
$\Gamma _{\rm A}$, ${\rm A}=1, \cdots ,
6$ is represented by $\sqrt 2 \psi _{\rm A}$, while the overall factor of
$\Gamma _7$ acts as a fermion number operator and flips the
worldsheet fermion boundary conditions to periodic ones. Putting all
together, we obtain
$$
\eqalign{
W_\Im
 &= -{i{\cal E}\over 64}
    \int_{-1}^1 \!\!\! d\alpha ~\int_0^\infty \!\!\! dT \,
\int {\cal D}p{\cal D}x \int\limits_{\rm P}{\cal D}\psi \; {\rm tr}\;
\chi \,
[\omega (p,{\cal A},{\cal H},{\cal K},\sqrt{2}\psi_a)\sqrt{2}\psi_6]( \tau =
0)
\cr
&
\qquad
\times{\cal P}\exp\biggl{\{ }\int_0^Td\tau\Bigl[ i\dot{x}_\mu p_\mu
- {1\over2}\psi_{\rm A}\dot{\psi}_{\rm A}
- {{\cal E}\over 2}
{\tilde\Sigma}^2_{(\alpha)}
(p,{\cal A},{\cal H},{\cal K},\sqrt{2}\psi_a)\Bigr]\biggr{\} } \; .
\cr}
\eqno (4.13)
$$
The boundary conditions on the worldline loop are periodic (P)
for both $x$ and
$\psi_{\rm A}$.
The $\psi_6$ insertion may be integrated since it is absent from
$\tilde\Sigma_{(\alpha)}^2$ and $\omega$; this allows us to change
$\psi_{\rm A}\dot{\psi}_{\rm A} \rightarrow \psi_a\dot{\psi}_a$ and view the
measure as
${\cal D}\psi = {\cal D}\psi_\mu{\cal D}\psi_5$.
Now by shifting the momentum integration as in the Sect. 3, we obtain
$$
W_{\Im} = -{i{\cal E}\over 64}
\int_{-1}^1 d\alpha \int_0^{\infty}{dT}\, {\cal N}\int\limits_{{\rm
P}}
{\cal D}x {\cal D}\psi \, {\rm tr} \, \chi
\bar{\omega} (0) \,
{\cal P} e^{-\int_0^Td\tau{\cal L}_{(\alpha)}(\tau )} \; .
\eqno(4.14)
$$
The worldline Lagrangian, ${\cal L}_{(\alpha)}(\tau )$, is given in
terms
of the worldline Lagrangian of the previous Section, ${\cal L} (\tau  )$, by
$$
\eqalignno{
{\cal L}_{(\alpha)}(\tau ) &= {\cal L}(\tau )
\big \vert_{\Phi  \rightarrow \alpha \Phi ~,~ B  \rightarrow \alpha B
{}~,~
                      K \rightarrow \alpha K } \; .
& (4.15)
\cr}
$$
The insertion operator, $\bar{\omega}$, is given by
$$
\bar{\omega}  = \sqrt{2} \,
\omega \Bigl( p_\mu \rightarrow \bigl[ {i\dot{x}_\mu \over {\cal E}}
+
(\alpha_+{\cal A}_\mu  + \alpha_-{\cal A}^c_\mu)
+ i\Gamma_\nu\Gamma_5 (\alpha_+{\cal K}_{\mu\nu}
+ \alpha_-{\cal K}^c_{\mu\nu})
\bigr] , \;
{\cal A}, \; {\cal H}, \; {\cal K}, \; \Gamma_a \Bigr)
\eqno (4.16a)
$$
followed by ${\Gamma_a \rightarrow \sqrt{2}\psi_a}$.  This works out to be
\footnote{${}^\dagger$} {In the insertion, the shift of the momentum by
non-commuting background fields must always be performed under the
anti-commutator.}
$$
\eqalignno{
\bar{\omega}
&= 2\psi_\mu \Bigl (-{2i\dot{x}_\mu \over {\cal E}}({\cal H} - {\cal H}^c)
			-{2\dot{x}_\nu\over {\cal E}}
		({\cal K}_{\mu\nu} - {\cal K}_{\mu\nu}^c)
                     - \{ ({\cal A}_\mu - {\cal A}_\mu^c),
			(\alpha_+{\cal H} + \alpha_-{\cal H}^c)  \}
		 \Bigr  )
&\cr
& + 2\psi_5   \Bigl( -i\partial_\mu ({\cal A}_\mu - {\cal A}_\mu^c)
			+ [{\cal A}_\mu , {\cal A}_\mu^c]
                     +  [{\cal H} , {\cal H}^c ]
                   + {3\over 2} [{\cal K}_{\mu\nu},{\cal K}^c_{\mu\nu}] \Bigr)
&\cr
&		+ 2 \psi_\mu\psi_\nu\psi_\rho \Bigl (
		 {\cal D}_\mu {\cal K}^c_{\nu\rho}
                     - {\cal D}^c_\mu {\cal K}_{\nu\rho}
                 + 2i \{ ({\cal A}_\mu - {\cal A}_\mu^c) ,
     (\alpha_+{\cal K}_{\nu\rho} + \alpha_-{\cal K}_{\nu\rho}^c)\} \Bigr )
&\cr
&
+ 2\psi_\mu\psi_\nu\psi_5  \Bigl ( {4i\dot{x}_\mu\over {\cal E}}
			({\cal A}_{\nu} - {\cal A}_{\nu}^c)
		+  i [{\cal H},{\cal K}^c_{\mu\nu} ]
             - i [{\cal H}^c,{\cal K}_{\mu\nu}]
             -4  \{ {\cal K}_{\mu\rho},{\cal K}^c_{\rho\nu} \}
& \cr
&
	 - 2i\{  ({\cal H} - {\cal H}^c),
		(\alpha_+{\cal K}_{\mu\nu} + \alpha_-{\cal K}_{\mu\nu}^c) \}
		- \psi_\rho\psi_\sigma
                     [{\cal K}_{\mu\nu},{\cal K}^c_{\rho\sigma}] \Bigr )  \; ,
& (4.16b) \cr}
$$
with
$$
\alpha_{\pm} \equiv { 1 \pm \alpha \over 2 } \; .
\eqno (4.17)
$$
It is clear that the
introduction of the parameter $\alpha$ breaks
the manifest chiral invariance, but it seems that this is the price to
pay for   obtaining a well-defined heat-kernel representation of
$W_\Im$.

In [7], we discussed the perturbation expansion of (4.14), and
we quote some of the result here.  Formulas for the bosonic sector are
given   exactly as in (3.17).  Writing $\psi_{\rm A} (\tau ) =
\psi^\prime_{\rm A}  (\tau   ) +
\psi_{\rm A} ^o$ where  $\dot \psi_{\rm A} ^o = 0$ and $\int_0^T d\tau
\psi^\prime_{\rm A} (\tau ) = 0$, the worldline fermion propagator is
$$
\langle\psi^\prime_{{\rm A}_1}(\tau_1) \psi^\prime_{{\rm A}_2}(\tau_2)
\rangle = {1 \over 2} \delta _{{\rm A}_1 {\rm A}_2}
\biggl ( {\rm sign}(\tau_1 -
\tau_2) -
{(\tau_1 - \tau_2)\over T} \biggr )\; .
\eqno (4.18)
$$
The integration over ${\cal D}\psi^\prime$ produces an extra overall
normalization factor of $-1$.
The integration over the worldline fermion zero modes is given by
$$
\int d^6\psi \; \psi_\mu
^o\psi_\nu^o\psi_\rho^o\psi_\sigma ^o\psi_5 ^o\psi_6 ^o  =
\varepsilon_{\mu\nu\rho\sigma} \; ,
\eqno (4.19)
$$
where our Euclidean space convention for the Levi-Civita tensor is
$\varepsilon_{1234} = 1$.

\vskip 0.3in

\noindent
{\bf 5. Worldline Treatment for the Internal Degrees of Freedom}

\bigskip

To complete our analysis, we derive the worldline formulation that
also promotes the internal degrees of freedom into Grassmann-valued
integrals. This representation has already been discussed extensively
in the   literature [1], but we shall give an independent derivation
of it here. We begin by   showing that the trace of the path ordered
exponential over an $n\times n$   Hermitian traceless
matrix $M(\tau)$,
$$
z[M] \equiv {\rm tr} \, {\cal P} e^{i\int_0^T d\tau M(\tau ) } \; ,
\eqno (5.1)
$$
may be re-expressed as a worldline path integral
$$
z[M]   = \Bigl( {\pi \over T}\Bigr)^n \sum _\varphi
       \int\limits_{\rm AP} {\cal D}\lambda^\dagger {\cal D}\lambda
{}~
       \epsilon ^{i\varphi
       (\lambda^\dagger \lambda + {n\over 2} - 1)}
       \, e^{-\int_0^T d\tau \bigl[ \lambda^\dagger \dot{\lambda}
       - i\lambda^\dagger M \lambda  \bigr] } \; .
\eqno (5.2)
$$
The summation symbol $\sum _\varphi$ may either be taken to be the
integral over the angle $\varphi$ from 0 to $2 \pi$, or, in the case
of $n\times n$ matrices, the sum over the values $2\pi k/n$ with
$k=1,\cdots,n$, in either case, suitably normalized to unity. The
$\varphi$-integration (or sum), must be included to properly project
the intermediate states in the path integral on coherent states of
occupation number 1. (When the discrete sum is used, this projection
is analogous to the GSO projection in string theory.)  The operator
$\lambda^\dagger\lambda$ in the exponent of (5.2) may be inserted at any
arbitrary value of $\tau$ which we shall take to be $\tau = 0$.

To show (5.1), we may perform the integration over $\lambda$ and
$\lambda^\dagger$ explicitly and obtain:
$$
z[M] =  \Bigl( {\pi \over T}\Bigr)^n
        \sum _\varphi \,
        {\rm Det} \Bigl[ {d\over d\tau} - M - {i\varphi \over T}
\Bigr]
        e^{i\varphi ({n\over 2} - 1)} \; .
\eqno (5.3)
$$
We evaluate the functional determinant in (5.3) as the infinite
product of the
eigenvalues of the operator $ {d\over d\tau} - M - {i\varphi \over T}$
with   anti-periodic boundary conditions on the eigenfunctions. This
leads   to
$$
z[M] = \Bigl( {\pi \over T}\Bigr)^n \, {T\over 2 \pi}
       \sum _\varphi  \, \prod_{k=1}^n \prod_{l=1}^\infty
       \, \biggl[ 1 - { (m_k + \varphi )^2 \over \pi^2(2l - 1)^2 }
\biggr]
       \, e^{i\varphi ({n\over 2} - 1)} \; ,
\eqno (5.4)
$$
where the $m_k$ are defined by
$$
{\cal P} e^{i\int_0^T d\tau M(\tau ) } = e^{i{\rm
Diag}(m_1,...,m_n)}   ~,
\qquad \quad
\sum _{k=1}^n m_k =0 \; .
\eqno (5.5)
$$
Finally, using the product formula for the cosine function,
we easily obtain the desired result,
$$
z[M] = \sum _\varphi \, \prod_{k=1}^n
       \bigl[ 1 + e^{i\varphi}e^{im_k} \bigr] e^{-i\varphi}
     =  {\rm tr} \, {\cal P} e^{i\int_0^T d\tau M(\tau ) } \; .
\eqno (5.6)
$$
Now, we could have perfectly well used periodic boundary conditions
for the integration over the worldline fermions $\lambda$ and
$\lambda^\dagger$ in (5.2).  In this case, the path integral (5.2)
would reduce to the  path ordered exponential (5.1) provided we replace
the normalization   of $(\pi/T)^n$ in (5.2) by $-(2\pi^2)^n$.

If there is an insertion in the trace, our method easily
generalizes to give
$$
  {\rm tr} \, \omega ( 0) \,
                {\cal P} e^{i\int_0^T d\tau M(\tau ) }
    =  N_{\rm BC} \sum _\varphi
       \int\limits_{\rm BC} {\cal D}\lambda^\dagger {\cal D}\lambda
       \,  \bigl \{ \lambda^\dagger \omega \lambda  ( 0) \bigr \}
       e^{i \varphi
       (\lambda^\dagger \lambda + {n\over 2} - 1)}
       \, e^{-\int_0^T d\tau \bigl[ \lambda^\dagger \dot{\lambda}
       -i \lambda^\dagger M \lambda  \bigr] } \; ,
\eqno (5.7)
$$
where $N_{\rm BC} = (\pi/T)^n$ or $-(2\pi^2)^n$ depending on whether
the
boundary conditions
for $\lambda$ and $\lambda^\dagger$ are anti-periodic or periodic,
respectively.

We may now recast our previous results for the path integral
reformulation
of the one-loop effective action where the trace over internal
degrees of
freedom
shall be represented by a worldline path integral over
a multiplet of $2N$ worldline fermions,
$\lambda$.
Combining the real and imaginary parts of the effective action
obtained in the previous two Sections with a sum over the boundary
conditions
(BC) of the worldline fermions, namely periodic (P) or
anti-periodic (AP)
boundary conditions on the worldloop gives
$$
\eqalign{
W[\Phi,\Pi,A,B,K]
&={1 \over 8}
  \int_0^{\infty} {dT \over T} \, {\cal N}
  \int {\cal D}x \; \sum_{\rm BC}
  \; \int\limits_{{\rm BC}} {\cal D}\psi{\cal D}\lambda^\dagger{\cal
D}\lambda
\cr
&\times \sum _\varphi e^{i\varphi (\lambda ^\dagger \lambda + N -1)}
  {\cal I}_{\rm BC}  (0)
 \, e^{-\int_0^Td\tau
 \bigl[{\cal L}_K(s,\tau )
 + \lambda^\dagger{\cal L}_I (\alpha,\tau)\lambda \bigr]} \; ,
\cr}
\eqno (5.8)
$$
with the worldline insertions given by
$$
\eqalignno{
{\cal I}_{\rm AP} ( 0)
&= \Bigl({\pi \over T}\Bigr)^{2N} \, \delta _{\alpha ,1}
   \; ,
&(5.9a)
\cr
\noalign{\hbox{and}}
{\cal I}_{\rm P} ( 0)
&=
-(2\pi^2)^{2N} \, {{\cal E}T\over 8} \int_{-1}^1 d\alpha
\lambda ^\dagger \chi \bar{\omega}  \lambda (0) \; .
&(5.9b)
\cr}
$$
Recall that $\bar{\omega} (\tau = 0)$
is the previous worldline insertion defined
in $(4.16)$ and $\chi$ is the fermion number defined in (4.6).
The worldline Lagrangians are
$$
\eqalignno{
{\cal L}_K (s,\tau )
&= {\dot{x}^2 \over 2{\cal E}} + {1\over 2}\psi_\mu\dot{\psi}_\mu
+ {1\over 2}\psi_5\dot{\psi}_5
+ \lambda^\dagger \dot{\lambda} \; ,
&(5.10a)
\cr
\noalign{\hbox{and}}
{\cal L}_I (\alpha,\tau )
&=\Bigr[  - i\dot{x}_\mu {\cal A}_\mu
          + {{\cal E}\over 2}{\cal H}^2
          - {{\cal E} \over 4}{\cal K}_{\mu\nu}{\cal K}_{\mu\nu}
          + i\psi_\mu\psi_5 \bigl (
              {\cal E}{\cal D}_{\mu}{\cal H}
              +2i {\dot x}_\nu {\cal K}_{\mu\nu} \bigr )
&  \cr
& \;\;\;\;\;\;
          + {i {\cal E} \over 2} \psi_\mu \psi_\nu \bigl (
              {\cal F}_{\mu\nu} +
            \{ {\cal H},{\cal K}_{\mu\nu}  \} \bigr )
& (5.10b) \cr
& \;\;\;\;\;\;
             - {\cal E}\psi_{\mu}\psi_\nu\psi_\rho \bigl ( \psi_5{\cal
             D}_\mu {\cal K}_{\nu \rho}
         +\half \psi_\sigma {\cal K}_{\mu\nu} {\cal K}_{\rho\sigma}
\bigr )
\Bigr]_{\Phi \rightarrow \alpha\Phi ~,~ B \rightarrow \alpha B ~,~
        K \rightarrow \alpha K} \; .
& \cr}
$$
Letting $\lambda$ transform under the representation
$(T_L \otimes 1) \oplus (1 \otimes T_R)$ of the chiral gauge group
$G_L \times G_R$, it is clear that for $\alpha = 1$ these Lagrangians are
manifestly $G_L \times G_R$ invariant.

\vskip 0.3in

\noindent {\bf Acknowledgments}

\bigskip

We gratefully acknowledge helpful conversations with Zvi Bern and
Daniel Cangemi on
worldline path integrals.

\bigskip

\noindent {\bf Appendix A : Alternative Worldline Path Integral
Reformulation}

\bigskip

The path integral reformulation of the effective action can be
carried
out in a way that does not exhibit chiral symmetry manifestly but
makes the
coupling to the background fields, $\Phi$, $\Pi$, $A$, $B$ and  $K$,
more explicit.  We are again interested in
the field theory defined classically by (2.1).  Then from (3.1) and
(4.1),
we have for the Euclidean one-loop effective action
$$
W[\Phi,\Pi,A,B,K] = -{1\over 2}\ln      {\rm Det}[\Sigma ]
                        -{i\over 2}{\rm arg}{\rm Det}[\Omega ] \; ,
\eqno (A.1)
$$
where explicitly, $\Sigma$ and $\Omega$ are given by (3.4) and (4.2)
respectively.
Regularizing the real and imaginary parts of $(A.1)$ directly as in
[7], gives
$$
W =  {1\over 4}\int_0^{\infty}{dT \over T} {\rm Tr}\, \biggl{\{ }
     e^{-{{\cal E}\over 2}T \Sigma ^2} -{{\cal E}T\over 8}
 \int^1_{-1} d\alpha \,
     (\Omega^2 - \Omega^{\dagger 2}) \,
     e^{-{{\cal E}\over 2}T
     \bigl[{1 \over 4}(\Omega + \Omega^{\dagger})^2
     +{\alpha \over 2}[\Omega, \Omega ^\dagger]
     -{\alpha ^2\over 4} ( \Omega - \Omega ^\dagger)^2 \bigr]}
     \biggr{\} } \; .
\eqno (A.2)
$$
By introducing a multiplet of $N$ worldline fermions,
the path integral reformulation [7] of $(A.2)$ may be expressed with
a sum over the boundary conditions (BC) of the worldline fermions,
namely periodic (P) or anti-periodic boundary conditions (AP) on the
worldloop:
$$
W
={1 \over 4} \int_{-1}^1 d\alpha
  \int_0^{\infty} {dT \over T} \, {\cal N}
  \int {\cal D}x \;
  \sum_{\rm BC} \; \int\limits_{{\rm BC}} {\cal D}\psi
  {\cal D}\lambda^\dagger {\cal D}\lambda \sum_\varphi
  e^{i\varphi (\lambda^\dagger \lambda + {N\over 2} - 1)}
   {\cal J}_{\rm BC}  ( 0)
  \, e^{-\int_0^Td\tau{\cal L}_{\alpha}(\tau )} \; ,
\eqno (A.3)
$$
with the worldline insertions given by
$$
\eqalignno{
{\cal J}_{\rm AP}(0)
&
= \Bigl({\pi\over T}\Bigr)^N \, \delta (\alpha - 1)
     \; , & (A.4a) \cr
\noalign{\hbox{and}}
{\cal J}_{\rm P} ( 0)
& = { iT \over 2 } \, (2\pi^2)^{N} \,
\lambda ^\dagger \Bigl [ \psi_\mu\psi_6 \bigl (
        2{\cal E} D_\nu K_{\mu\nu}  - 2i \dot{x}_\mu\Phi
                      - i{\cal E}  [ \Pi , B_\mu ]  )
    + {\cal E} \psi_5 \psi_6 \bigl (\{ \Phi , \Pi \} - D_\mu B_\mu \bigr )
& \cr
& \qquad +	\psi_\mu\psi_\nu{\cal E}(2i\alpha[\Phi,K_{\mu\nu}]
	+ 4\alpha\{ K_{\mu\rho},K_{\rho\nu} \})
	+ \psi_\mu\psi_5{\cal E}(i\alpha[\Phi,B_\mu] +
		2\alpha \{ B_\nu, K_{\mu\nu} \})
& \cr
& \qquad +\psi_\mu\psi_\nu\psi_\rho \bigl (4\dot{x}_\mu \psi_6 K_{\nu\rho}
	+ 4\alpha{\cal E}\psi_\sigma[K_{\mu\nu},K_{\rho\sigma}]
	- 2\alpha{\cal E}\psi_5[B_\mu,K_{\nu\rho}] \bigr )
& \cr
& \qquad
   +\psi_\mu\psi_\nu\psi_5\psi_6 \bigl ( 4\dot{x}_\mu B_\nu
              + 2i{\cal E} \{ \Pi , K_{\mu\nu} \}\bigr )
\Bigr ] \lambda ( 0 ) \; , & (A.4b) \cr
\noalign{\hbox{and the worldline Lagrangian given by}}
{\cal L}_\alpha (\tau )
&
={ \dot{x}^2 \over 2{\cal E} } + {1\over 2}\psi_\mu\dot{\psi}_\mu
+ {1\over 2}\psi_5\dot{\psi}_5 + {1\over 2}\psi_6\dot{\psi}_6
+ \lambda^\dagger \dot{\lambda}
& \cr
&\quad
+\lambda ^\dagger \Bigl [
   -i\dot{x}_\mu A_\mu
   + { {\cal E}\over 2 } \Pi^2
   +{ {\cal E}\over 2 } \alpha^2 \Phi^2
   -   {\cal E}          \alpha^2 K_{\mu\nu}K_{\mu\nu}
   + {\cal E}\psi_\mu\psi_5 \bigl (iD_\mu \Pi
   + i\alpha^2 \{ \Phi ,B_\mu    \} \bigr )
& \cr
& \qquad \quad
   + \alpha \psi_\mu\psi_6 \bigl ( i{\cal E} D_\mu \Phi
   -4 \dot{x}_\nu K_{\mu\nu} -i{\cal E} \{ \Pi ,B_\mu \} \bigr )
+ \alpha\psi_5\psi_6
\bigl (2\dot{x}_\mu B_\mu  - {\cal E}[\Phi , \Pi ] \bigr )
& \cr
& \qquad \quad
 +{ {\cal E} \over 2 } \psi_\mu \psi_\nu \bigl (i F_{\mu\nu}
   +\alpha^2  [B_\mu ,B_\nu ]
   + 2i\alpha^2 \{ \Phi ,K_{\mu\nu} \} \bigr )
& \cr
& \qquad \quad
-2{\cal E}\alpha\psi_\mu\psi_\nu\psi_5\psi_6 \bigl (
     D_\mu B_\nu  -i [\Pi ,K_{\mu\nu}] \bigr )
-2{\cal E}\alpha\psi_\mu\psi_\nu\psi_\rho\psi_6D_\mu K_{\nu\rho}
& \cr
& \qquad \quad
- 2{\cal E}\alpha^2\psi_\mu\psi_\nu\psi_\rho\psi_5\{ B_\mu
,K_{\nu\rho}\}
-  {\cal E}\alpha^2\psi_\mu\psi_\nu\psi_\rho\psi_\sigma
   \{ K_{\mu\nu} ,  K_{\rho\sigma} \} \Bigr ] \lambda \; , & \cr
& & (A.5) \cr}
$$
where the covariant derivatives are now taken only with respect to
the vector gauge field $A_\mu$, and $F_{\mu \nu}$ is the vector field
strength
$$
D_\mu  = \partial _\mu  - iA_\mu  ~,
\qquad
  F_{\mu\nu} = \partial _\mu A_\nu - \partial _\nu A_\mu - i
  [A_\mu, A_\nu] \; .
\eqno (A.6)
$$

We have checked that the imaginary part of $(A.3)$ correctly reproduces
the
chiral anomaly in four
dimensions.  We have also checked that our Lagrangian $(A.4b)$ matches term
by term with the result proposed in [8] in the abelian limit (with $K = 0$).
We unfortunately were unable to determine whether or not our result for the
antiysmmetric tensor couplings in $(A.4b)$ agrees with the result presented in
[9].

\bigskip

\noindent
{\bf References}

\bigskip

\item{[1]} R.P. Feynman, Phys. Rev. {\bf 80} (1950) 440; R.P.
Feynman, Phys.  Rev. {\bf 84} (1951) 108; J. Schwinger, Phys. Rev.
{\bf 82}   (1951) 664; E. S. Abers and B. W. Lee, Phys. Rep. {\bf 9C}
(1973)   1; R. Casalbuoni, J. Gomis and G. Longhi, Nouvo Cimento {\bf
24A} (1974) 249; F. A. Berezin and M. S. Marinov, JETP Lett. {\bf 21}
(1975) 320; R. Casalbuoni, Nuovo Cimento {\bf 33A} (1976)
389;  L. Brink, P. DiVecchia and P. Howe, Nucl. Phys. {\bf B118}
(1977) 76; A.P. Balachandran, P. Salomson, B. Skagerstam and J.
Winnberg, Phys. Rev. {\bf D15} (1977) 2308; A. Barducci, R. Casalbuoni and
L. Lusanna,  Nucl.
Phys. {\bf B124} (1977) 93; M.B. Halpern, P. Senjanovic and
A. Jevicki, Phys. Rev. {\bf D16} (1977) 2476; M.B. Halpern
and  W. Siegel, Phys. Rev. {\bf D16} (1977) 2486; A. Barducci, F. Bordi and
R. Casalbuoni,
Nuovo Cimento {\bf 64B} (1981) 287; D. Cangemi, E. D'Hoker and G. Dunne,
Phys. Rev {\bf D51} (1995) R2513; A. G. Morgan,
Phys. Lett. {\bf B351} (1995) 249.

\item{[2]} A.M. Polyakov, {\it Gauge Fields and Strings}, Harwood
Academic  Pub., 1987.

\item{[3]}S. Cecotti and L. Girardello, Phys. Lett. {\bf 110B} (1982)
39; L. Alvarez-Gaum\'e, Commun. Math. Phys. {\bf 90} (1983)
161; L. Alvarez-Gaum\'e and E. Witten, Nucl. Phys. {\bf B234}
(1983) 269; D. Friedan and P. Windey, Nucl. Phys. {\bf B235} (1984) 395.

\item{[4]}Z. Bern and D.A. Kosower, Phys. Rev. {\bf D38} (1988) 1888;
Z. Bern and D.A. Kosower, Phys. Rev. Lett. {\bf 66}   (1991) 1669; Z.
Bern and D.A. Kosower, Nucl. Phys. {\bf B379}   (1992) 451.

\item{[5]}M.J. Strassler, Nucl. Phys. {\bf B385} (1992) 145;
D. G. C. McKeon, Ann. Phys. {\bf 224} (1993) 139; D. G. C.
McKeon and A. Rebhan, Phys. Rev. {\bf D48} (1993) 2891.

\item{[6]}M. G. Schmidt and C. Schubert, Phys. Lett. {\bf 318} (1993)
438; M. G. Schmidt and C. Schubert, Phys. Lett. {\bf 331} (1994) 69.
M.   Mondrag\'on,  L. Nellen, M.G. Schmidt and C. Schubert, Phys. Lett.
          {\bf 351B} (1995) 200.

\item{[7]} E. D'Hoker and D. G. Gagn\'e, Preprint UCLA/95/TEP/22
          (hep-th/9508131), UCLA, 1995.

\item{[8]} M. Mondrag\'on, L. Nellen, M.G. Schmidt and C. Schubert,
Preprint IASSNA-HEP-95/74, HD-THHEP-95-43, HUB-EP-95/17
(hep-th/9510036), Heidelberg, 1995.

\item{[9]} J.W. van Holten,  Preprint NIKHEF-95-055
(hep-th/9510021), NIKHEF, 1995.

\item{[10]}M.B. Green, J.H. Schwarz and E. Witten, {\it Superstring
Theory}, Cambridge University Press, 1987; E. D'Hoker and D.H.
Phong, Rev. Mod. Phys. {\bf 60} (1988) 917.

\end